\documentclass{emulateapj}

\newcommand{\beq}{\begin{equation}}
\newcommand{\eeq}{\end{equation}}
\newcommand{\beqar}{\begin{eqnarray}}
\newcommand{\eeqar}{\end{eqnarray}}

\shorttitle{$1/f$ noise and asymmetric beam impact on {\sc{Planck}} NG searches}
\shortauthors{Donzelli et al.}

\begin{document}

\title{Impact of the $1/f$ noise and the asymmetric beam on non-Gaussianity searches with {\sc{Planck}}}

\author{Simona Donzelli\altaffilmark{$\star$,1,2,4}, Frode K. Hansen\altaffilmark{1,2}, Michele Liguori\altaffilmark{3}, and Davide Maino\altaffilmark{4}}

\altaffiltext{$\star$}{simona.donzelli@astro.uio.no}
\altaffiltext{1}{Institute of Theoretical Astrophysics, University of Oslo, P.O. Box 1029 Blindern, N-0315 Oslo, Norway}
\altaffiltext{2}{Centre of Mathematics for Applications, University of Oslo, P.O. Box 1053 Blindern, N-0316 Oslo, Norway}
\altaffiltext{3}{Department of Applied Mathematics and Theoretical Physics, Centre for Mathematical Sciences, University of Cambridge, Wilberfoce Road, Cambridge, CB3 0WA, United Kingdom}
\altaffiltext{4}{Dipartimento di Fisica, Universit\`a degli Studi di Milano, via Celoria 16, 20133 Milano, Italy}

\begin{abstract}
We study the impact of correlated instrumental noise 
and non-circular antenna beam patterns on primordial non--Gaussianity 
analysis. The two systematic effects
are reproduced in the case of the {\sc{Planck}} mission,
using {\sc{Planck}}-like realistic simulations.
The non--Gaussian analysis is
conducted with different approaches.  First we adopt a blind approach, using
the Spherical Mexican Hat Wavelet and the Minkowski functionals, and then
a $f_{\rm NL}$ estimator. We look respectively for false primordial non--Gaussian 
detections and for bias or variance increase in the estimated $f_{\rm NL}$ value.
Even if some slight effects are present, we can not observe any significant
impact of the $1/f$ noise and the asymmetric beam on non--Gaussianity searches in the 
context of the {\sc{Planck}} mission.
\end{abstract}


\section{Introduction}

Today  the Gaussianity of the primordial 
fluctuations has become a key observable in cosmology. 
The standard single-field inflationary 
models predict 
undetectable deviations from Gaussianity \citep{Acquaviva03,Maldacena03}, while 
alternative scenarios
can produce different levels of non-Gaussianity,
 usually parametrized by the 
non--linearity parameter $f_{\rm NL}$
\citep{KomSperg01,BartoloReview,lindemu97,lyth03,babich04,
chen07a,holmtol08,chen07b,langl08}.
Since different models predict different limits
on the $f_{\rm NL}$ parameter,
the constrains on 
Gaussianity can discriminate between different scenarios.
Nevertheless testing non-Gaussianity is challenging. 
Firstly because the primordial deviations from Gaussianity
are expected to be tiny.
Secondly, residual foregrounds and systematic effects could 
introduce non-Gaussian signatures which can mimic the primordial 
non-Gaussianity signal.
Moreover it is interesting to look
for any non--Gaussian signal, not only of 
the kind predicted by Inflationary models, but
it does not exist a statistical test sensitive to all 
possible non-Gaussian manifestations
of a random field. Therefore the problem must be approached with qualitatively 
different tests. The actual and forthcoming 
Cosmic Microwave Background (CMB) experiments have the capability to strongly
constrain primordial non Gaussianity, thanks to their 
high sensitivity and angular resolution.
In order to achieve this goal an accurate control on the systematic artifacts 
is essential. Today the best CMB data come from the 
Wilkinson Microwave Anisotropy Probe ($WMAP$) satellite. 
The $WMAP$ team analysis on the 5 year data found them consistent 
with Gaussianity, putting 
the following constrains on the non-linearity parameter 
$f_{\rm NL}$: $-9<f_{\rm NL}^{local}<111$ 
(95\% CL) and $-151<f_{\rm NL}^{equil}<253$ (95\% CL) \citep{komatsu08}.
These parameters constrain non--Gaussian models which predict large contributions
to the bispectrum respectively from the squeezed and the 
equilateral configurations \citep{babich04,lindemu97,lyth03,babich04,
chen07a,holmtol08,chen07b,langl08}. 
These limits have been improved with an optimal analysis by
 \citet{smith09} and \citet{senatore09}, 
finding $-4<f_{\rm NL}^{local}<80$ (95\% CL) and $-125<f_{\rm NL}^{equil}<435$ 
(95\% CL) respectively.
Consistent results with different methods 
were found by other groups \citep{curto08,pietrobon08,
rudjord09}.
However other studies have shown some hints of non-Gaussian features at 
the $2-3\sigma$ level. These include the presence of a very cold spot in the data, 
deviations of the expected statistics of peaks from Gaussian expectations, 
non-Gaussian distribution of the anisotropies in the southern emisphere and others 
($e.g.$ \citep{vielva04,Hansen08,mcewen08}). Moreover a detection of 
a non-zero $f_{\rm NL}$ has been achieved by \citet{yadavwan08}
on the $WMAP$ 3 year dataset. With the
same statistical method adopted by the $WMAP$ team, they 
found $27<f_{\rm NL}^{local}<147$ (95\% CL).
The forthcoming data of European Space Agency (ESA) CMB mission {\sc Planck} will help 
to clarify these 
puzzles. {\sc Planck} will image the whole sky with outstanding sensitivity, angular resolution and  frequency coverage \citep{bluebook}. To take advantage of this top-level 
performances, a fine control of all possible systematic effects and of their
impact on the tests of non-Gaussianity  is crucial. 

In this paper we consider
two instrumental effects: the correlated noise, also known as $1/f$ noise, and the 
asymmetry of the antenna beam. We have investigated their impact on 
primordial non-Gaussianity searches 
carried out with different
statistical tests: Spherical Mexican Hat Wavelets, Minkowski functionals
and a $f_{\rm NL}$ estimator, often dubbed as ``KSW'' estimator \citep{KSW05}. The systematic effects are studied in the context of the {\sc Planck} 
mission.

The paper is organized as follows. In Section~\ref{syst.eff} we describe the main characteristics of the {\sc Planck} mission and
the systematic effects under study.
Then in Section~\ref{stat.tools} we shortly review the statistical tests used in this work.
The analysis method applied and the results obtained are presented in Section~\ref{analysis}.
Conclusions are drawn in Section~\ref{conclus}.

\section{The systematic effects}\label{syst.eff}
\subsection{The {\sc Planck} mission}

{\sc Planck}, successfully launched on may 14, 2009, 
consists of two different instruments, called the Low and the High Frequency 
Instrument (LFI and HFI respectively). LFI covers the frequency range $30$--$70$ GHz 
with an array of receivers based on High Electron Mobility Transistor (HEMT) 
amplifiers. The HFI receivers are instead based
on bolometers, and operate between $100$ and $857$ GHz. 
Each feed horn collects radiation from the telescope and feeds it into one 
or more detectors. The satellite observes the 
sky from the 2nd Earth-Sun Lagrangian point (L2), spinning at $\sim1$ rpm 
with the spin axis kept in anti-Sun direction. The telescope
field-of-view is offset from the spin axis by $\sim 85^\circ$, so that the observed
patch traces large circles on the sky. The detectors will make repeated
 observations of the same sky circles before repointing the spin axis.
As the spin axis follows the Sun, the observed circles
 sweep through the sky at a rate of $1^\circ$day$^{-1}$  
\citep{bluebook}.

In order to achieve the precision aimed in the {\sc Planck} CMB data, 
the increase in angular resolution and sensitivity must be 
supported by an excellent control of the systematic effects. 
In the following we describe two typical instrumental effects: the  $1/f$ noise and the
asymmetric beam.

\subsection{The $1/f$ noise}
HEMT amplifier gain fluctuations, noise temperature fluctuations, thermal instabilities 
produce low-frequency noise in the data streams, also called $1/f$ noise.
In general the instrumental noise can be described as a sum of white and 
correlated noise, writing its Power Spectral Density in the form:
\beq\label{PSDoof}
P_{noise}(f)=\left[1+\left(\frac{f_{k}}{f}\right)^\alpha\right]
\frac{\sigma^2}{f_s},
\eeq
where $f_{k}$ is the ``knee'' frequency at which the white and the $1/f$ noise 
contribution are 
equal, $f_s$ is the sampling frequency and $\sigma$ is the white noise standard 
deviation per sample integration time ($t=1/f_s$). The value of the 
spectral index $\alpha$ depends on the dominant physical process which 
generates low-frequency noise.
 If not corrected the $1/f$ noise leads, 
coupled with the observing strategy, to 
unwanted stripe-wise structures in the final sky map. 
The effect of the $1/f$ noise can be approximated by a constant offset for 
each scan circle \citep{janssen96}.
If the scanning strategy is redundant, as the {\sc Planck} strategy actually is,
it is possible to derive 
the amplitudes of these offsets considering the intersections between different scan
circles. This is the key feature of the destriping techniques for map-making
($e.g.$ \citet{maino02,ashdown07a,elina04}).

\begin{figure}[t]
\centering
\includegraphics[angle=90,width=6.5cm]{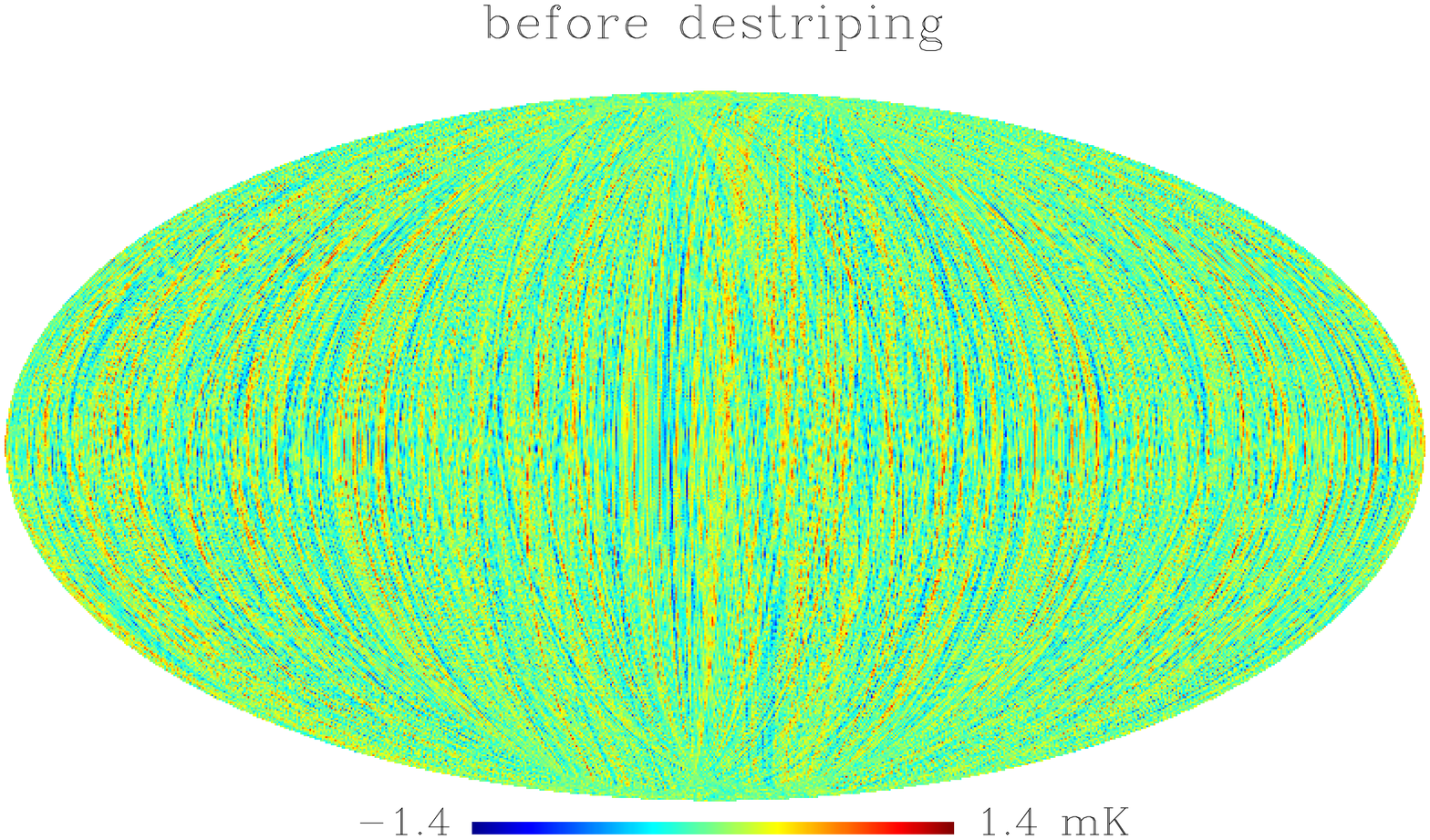}
\includegraphics[angle=90,width=6.5cm]{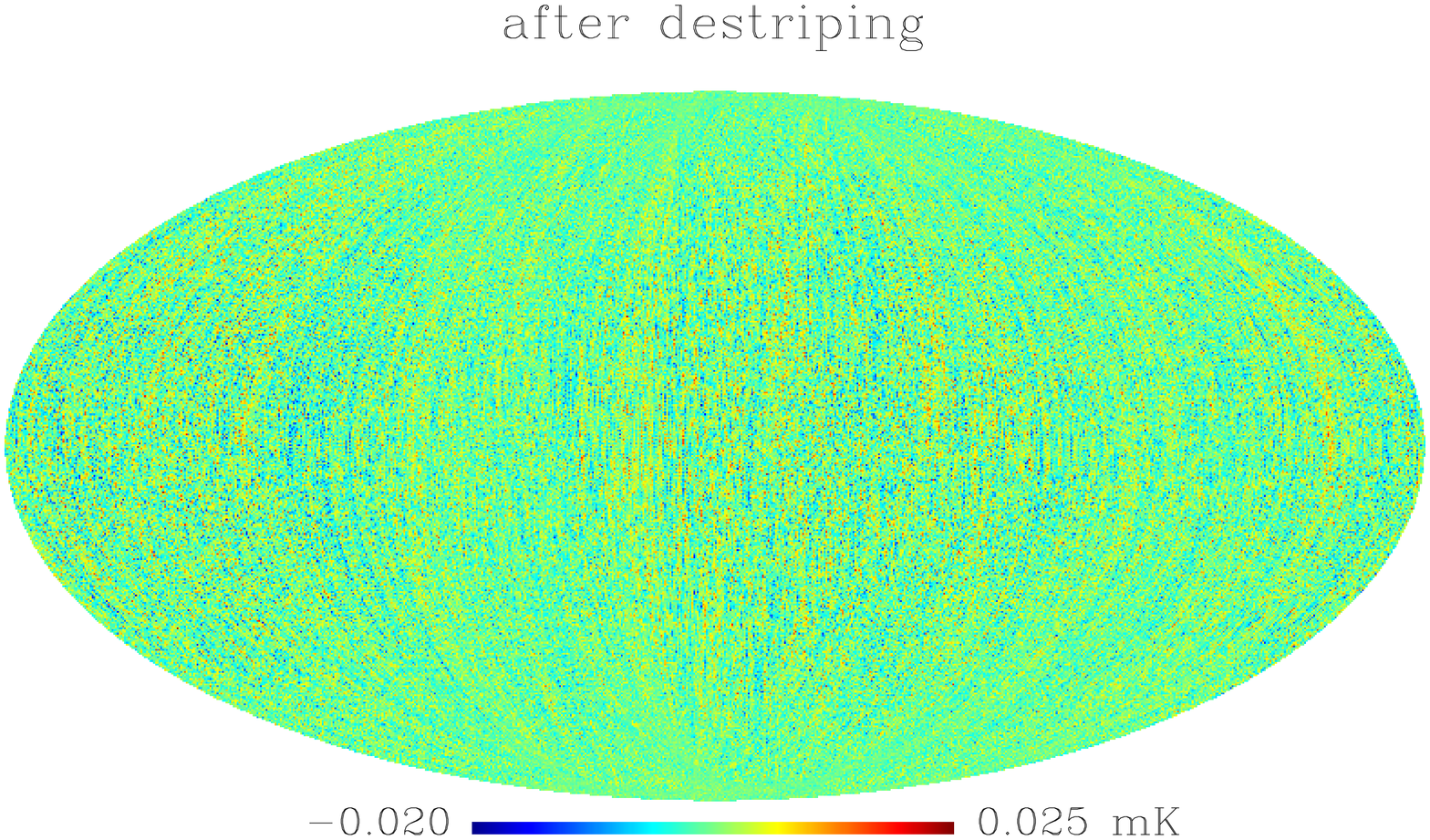}
\caption{(top) Map of white and $1/f$ noise before destriping; (bottom) after destriping stripes are visible due 
to residual $1/f$ noise (the white noise contribution has been subtracted). 
We remark the different peak-to-peak values of the maps.}
\label{mapdestrivsnodestri}
\end{figure}

\begin{figure}[t]
\centering
\includegraphics[angle=90,width=0.5\textwidth]{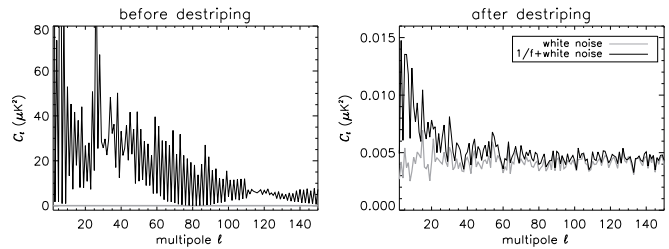}
\caption{The black line is the power spectrum of a map with white and $1/f$ noise  before (right) and  after (left) destriping (for $f_{k}=0.05$ Hz, $\alpha=1.7$). The gray line is the power spectrum of a map with only white noise. Note the different values range of the two plots.}
\label{cldestrivsnodestri}
\end{figure}

A destriping method can signiﬁcantly reduce  the $1/f$ noise.
Anyway a residual contribution is still present after the 
correction. There are residual stripes in the map  (Fig.~\ref{mapdestrivsnodestri}),
and in the angular power spectrum we can observe an excess of noise respect 
to the white noise level at low multipoles $\ell\lesssim 50$ 
(Fig.~\ref{cldestrivsnodestri}).
Analytical studies have shown that the residual $1/f$ noise introduces 
a block-diagonal structure in the covariance matrix of the harmonic coefficients
of the map. The off-diagonal terms have been found to be much smaller 
than the diagonal components, but not completely negligible 
in comparison to the tiny non-Gaussian signal we are looking for \citep{Efstathiou05}.
Therefore is important to investigate the impact of these correlations on the 
non-Gaussianity statistical tests.


\subsection{The asymmetric beam}

\begin{figure}[t]
\centering
\includegraphics[width=7cm]{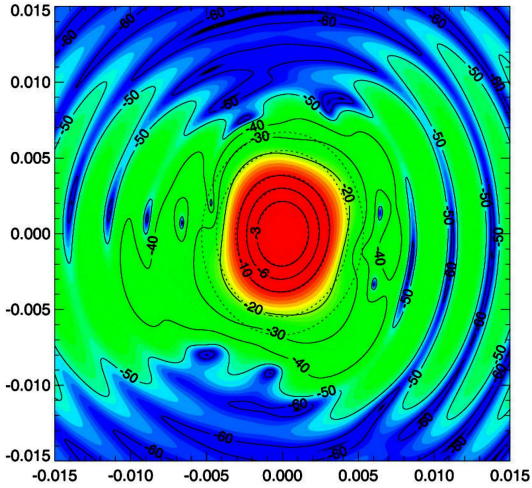}
\caption{Main beam contour plot of one of the {\sc{Planck-LFI}} 70 GHz feed horns
(simulated assuming an ideal telescope). 
The dotted lines are the bivariate Gaussian fitting 
(color scale: -90$\div$0 dB).}
\label{fig:main_beam}
\end{figure}

Actual CMB observation requires  multi-frequency focal plane arrays to achieve proper 
angular resolution, sensitivity and spectral coverage. This is indeed the case
of {\sc Planck}, where we have the HFI focal plane, with $36$ horns, inserted into the ring formed by the $11$ LFI horns. As a consequence, some feed 
horn can not be located close to the focal plane center, so that the beam pattern is asymmetric rather than circular.
Realistic {\sc Planck} optical simulations of the feed horns coupled to the telescope
\citep{sandri04,brossard04} have shown that the main beam patterns are well fitted 
by an elliptical shape given by a bivariate Gaussian (Fig.~\ref{fig:main_beam}).  
Moreover the finite 
response in time of the detectors leads to an elongation of the beam along 
the scan direction. The horns scan the sky mainly in the same direction, perpendicular 
to the ecliptic plane. Only the ecliptic poles are swept in several directions, 
which makes the effective beam more symmetric.

The convolution of the sky signal in a map $T(\hat r)$ with the beam  
can be written in harmonic space as
\beq\label{mapmak2}
T(\hat r)=\sum_\ell\sum_{m=-\ell}^{m=\ell} B_{\ell m}a_{\ell m}Y_{\ell m}(\hat r),
\eeq
where $Y_{\ell m}$ are the spherical harmonics and $B_{\ell m}$ and $a_{\ell m}$ are the 
harmonic expansions of the beam and the sky signal respectively.
For a circular symmetric beam we have  $B_{\ell m}=B_{\ell m'}$, but if the beam is asymmetric 
$B_{lm}$ is no longer isotropic and it depends  also on the 
azimuthal direction $m$ \citep{wu01}. The azimuthal dependence increases the 
deconvolution complexity and the computational cost, thus an exact deconvolution 
of an asymmetric beam is not straightforward. A beam not correctly measured and 
treated in the data analysis pipeline may artificially distort the underling structure
of the CMB signal, introducing a spurious dependence on the azimuthal direction and
messing up the  isotropy of the field.
We will investigate the impact of this effect in the non-Gaussianity searches
(see \citet{ashdown09} for a study of the asymmetric beam impact on map-making).

\section{Statistical tests of Non-Gaussianity}\label{stat.tools}

In this work we have made use of two qualitatively different 
statistical approaches
to test non-Gaussianity. 
In the first approach no assumption is made about the 
non--Gaussian model, and the Gaussian hypothesis is tested. In the 
following we will refer to such tests as blind tests.
In this work we have chosen 
to use the Spherical Mexican Hat Wavelet 
(SMHW) and the Minkowsky functionals (MF hereafter). 
This tests make use of Monte Carlo (MC) simulations to calibrate the Gaussian
behavior. A departure from the MC calibration is interpreted as a 
non-Gaussian detection. 
The blind approach has the advantage to be model independent, but it has a 
weak statistical power, $i.e.$ it is difficult to interpret
the meaning of an eventual detection of non-Gaussianity in the data.
In order to put quantitative constraints on the
non-Gaussianity level we have to adopt another approach.
Testing specific non--Gaussianity arising from inflationary models
we can put limits to the non-linearity parameter $f_{\rm NL}$.
We have used a cubic statistic
to estimate the $f_{\rm NL}$ parameter for both the local and the equilateral
non-Gaussianity. 

Given the difference in the two approaches,
also the possible impacts of the systematic effects are different.
For the blind tests the issue is in the MC calibration, since
it is not computationally manageable to include the $1/f$
noise or the asymmetric beam in
sets usually of thousands of simulations. Therefore the presence 
of the systematics in the data can in general cause departures from the
MC calibration, that can lead to false non-Gaussian detections.
Also in the non-blind cubic statistic it is often not feasible to take
exactly into account the systematic effects, due to the complexity of the 
numerical or analytical 
treatment involved. A not accurate handling of 
the systematics can potentially introduce a bias in the 
estimated  $f_{\rm NL}$. Even if on average the effect is zero, and there is no
bias, the spurious correlations introduced by the systematics can increase
the variance of the estimator. 
An example of this behavior is the effect of the
anisotropic noise, which introduces correlations between small and large scales.
If not corrected, these leads to a noticeable increase of the variance at high multipoles
\citep{creminelli06,smith09}.

Before analysing the impact of the systematics effects, we will briefly describe the 
statistical methods applied in the study.

\subsection{SMHW}
The main idea of the wavelet technique is to decompose 
the temperature field into a set of coefficients, which
contain information about the signal at a given position and scale. 
We can perform a continuous analysis, taking the scale and the position in a continuum 
of values. The wavelets are a convenient technique 
to study non–-Gaussianity, since the wavelet transform is linear. Thus the 
wavelet coefficients preserve the Gaussianity properties of the original field.
Moreover they allow us to study the non-Gaussian features at different scales.
In particular the SMHW enhances the non-Gaussian signatures with spherical symmetry.
The SMHW is the extension
to the sphere of the Euclidian Mexican Hat Wavelet \citep{Cayon00,martinez02}. 
It is given by
\beqar
\Psi (\theta; R)&=& \frac{1}{\sqrt{2\pi}R\,N}
\left[1+\left(\frac{y}{2}\right)^2\right]^2
\left[2 - \left(\frac{y}{R}\right)^2\right]e^{-\frac{y^2}{2R^2}},\\
N(R)&\equiv& \left(1 + \frac{R^2}{2} + \frac{R^4}{4}\right)^{1/2},\nonumber
\eeqar
where $R$ is the scale, $(\theta ,\phi )$ represent polar coordinates on the sphere,
$(y\equiv 2\tan \frac{\theta}{2}, \phi)$ are polar coordinates in the tangent plane to the North pole, and $N(R)$ is a normalisation constant. Given the temperature field on the 
sphere $T(\theta, \phi)$, the wavelet coefficients are defined as:
\beqar
&&w(\theta ,\phi ; R) = \int d\theta^{\prime}\,d\phi^{\prime}\sin \theta^{\prime}
\,\tilde{T}({\bf x} + {\bf \mu})\Psi (\theta^{\prime}; R),\,\\
&&{\bf x} = 2\tan \frac{\theta}{2}(\cos\phi,\sin\phi), \,
{\bf \mu}=2\tan \frac{\theta^{\prime}}{2}(\cos\phi^{\prime} ,\sin\phi^{\prime}).\nonumber
\eeqar
Several statistics of the wavelet coefficients $w_i$ can be studied at a 
given scale $R$. In this work we have analyzed two simple non-Gaussian estimators: the 
 skewness $S(R)$, $i.e.$ the degree of symmetry about the maximum,
and the kurtosis $K(R)$, which indicates the
degree to which a statistical frequency curve is peaked. Assuming
zero mean at each scale we have:
\beqar\label{skewkurt}
S(R)&=&\frac{1}{N_R}\sum_{i=1}^{N_R}\left(\frac{w_i(R)}{\sigma(R)}\right)^3\,,\\
K(R)&=&\frac{1}{N_R}\sum_{i=1}^{N_R}\left(\frac{w_i(R)}{\sigma(R)}\right)^4-3,\nonumber
\eeqar
where $\sigma^2(R)=\left(\sum_{i=1}^{N_R} w_i(R)^2\right)/N_R$ is the variance and $N_R$ is the number of coefficients (equal to the number of pixels $N_{pix}$ in our case).

\subsection{Minkowski functionals}

A general theorem of integral geometry states that the morphology of a 
convex body in $D$-dimensional space can be
completely described by $D+1$ Minkowski functionals \citep{minkowski1903}. 
In the case of the two-dimensional CMB field we consider the excursion sets $Q$, 
defined as the set consisting of the parts of the sky in which the temperature 
anisotropies amplitude $u$ is above a threshold 
$\nu$: $Q\equiv Q(\nu)=\{u| (u-\langle u \rangle)/\sigma_0>\nu\}$, where
$\sigma_0^2=\langle u^2 \rangle - \langle u \rangle^2$. 
Each of the functions has an explicit analytic expression 
for a Gaussian field \citep{tomita86}. Thus they provide an appropriate statistical
test for Gaussianity. 
Different algorithms have been implemented to estimate the MF
of a pixeled CMB sky map. In this work we have developed our algorithm 
following the 
approach of \citet{eriksen04}, using the derivative computation of the 
HEALPix\footnote{\tt{http://healpix.jpl.nasa.gov}} routines \citep{healpix}.
 The first functional, called
the area functional, is defined as:
\beq
\mathcal{A}(\nu) = \frac{1}{A_{obs}} \int_Q dA \approx
  \frac{N_{pix}(u >
\nu)}{N_{pix}^{\textrm{tot}}},
\eeq
and is equal to the total area of $Q$. The second equality holds in the HEALPix pixelization, in which all pixels have exactly equal area. 
Then we define the length functional, $i.e.$ the total length of the boundary $\partial Q$ of the excursion set:
\beq
\mathcal{L}(\nu) = \int_{\partial Q} d\ell
\eeq
where $\mathcal{L}$ is normalized to have a unit integral value: $ \int \mathcal{L}(\nu) d\nu = 1$.
Finally the third functional is the genus, which measures the connectivity of $Q$, being equal to the number of isolated regions (hot spots) minus the numbers of holes (cold spots). It can be defined by:
\beq
  \mathcal{G}(\nu) = \frac{N_{\rm{max}}(\nu) +
  N_{\rm{min}}(\nu) - N_{\rm{sad}}(\nu)}
{N_{\rm{max}}(-\infty) +
  N_{\rm{min}}(-\infty) + N_{\rm{sad}}(-\infty)},
\eeq
$i.e.$ as the number of maxima and minima minus the number of saddle
points in the excursion set,
divided by the sum of all stationary points in the whole
un-thresholded field.
In addition we use another morphological descriptor: the skeleton length, defined
as the zero-contour line of the map $\mathcal{S}$:
\beq\label{eq:S_sphere}
\mathcal{S} = T_{;\phi} T_{;\theta} (T_{;\phi\phi} -
T_{;\theta\theta}) + T_{;\phi\theta} (T_{;\theta}^2 -
T_{;\phi}^2),
\eeq
where semicolons denote the covariant derivatives. We compute the length of 
the skeleton of that part of the field that lies above some threshold $\nu$,
normalized by the length of the skeleton of the whole un-thresholded
field. The zero-contour lines of
$\mathcal{S}$ can be interpreted as the set of lines that extend from extremum to
extremum along lines of maximum or minimum gradient. For details of the 
MF algorithm see \citet{eriksen04}.

\subsection{$f_{\rm NL}$ estimator}\label{fnlestimator}
The three-point correlation, or the bispectrum, function is the 
lowest-order statistic able to distinguish non-Gaussian from Gaussian 
perturbations. Indeed for Gaussian CMB the expectation value is exactly zero.
Therefore we can look at the bispectrum as a natural observable to constrain 
the non--Gaussianity amplitude predicted by inflationary models,
determined by the $f_{\rm NL}$ parameter. After measuring the bispectrum modes 
of a given map we can build a cubic estimator of  $f_{\rm NL}$
in the following way:
\beq\label{est}
\hat{f}_{\rm NL}= \frac{{\mathcal S}_{\rm prim}}{N}.
\eeq
${\mathcal S}_{\rm prim}$  is a cubic statistic defined as
\beq\label{eq:Sprim}
{\mathcal S}_{\rm prim} = \sum_{\ell_1\le \ell_2\le \ell_3}
  \frac{\tilde{B}_{\ell_1\ell_2\ell_3}^{obs}\tilde{B}_{\ell_1\ell_2\ell_3}^{\rm prim}}
  {\tilde{C}_{\ell_1}\tilde{C}_{\ell_2}\tilde{C}_{\ell_3}},
\eeq
where $\tilde{B}_{\ell_1\ell_2\ell_3}^{obs}$ is the bispectrum extracted from the 
observed map and $\tilde{B}_{\ell_1\ell_2\ell_3}^{\rm prim}$ is the ansatz for $f_{\rm NL}=1$
 of the non-Gaussian model assumed, while $N$ is a normalisation factor 
defined as
\beq\label{N_Sprim}
N=\sum_{\ell_1\le \ell_2\le \ell_3}
  \frac{(\tilde{B}_{\ell_1\ell_2\ell_3}^{\rm prim})^2}
  {\tilde{C}_{\ell_1}\tilde{C}_{\ell_2}\tilde{C}_{\ell_3}},
\eeq
where $\tilde{C}_\ell$ is the theoretical power spectrum of the model, after including
the instrumental beam and noise. 

We adopt the fast numerical implementation of the estimator in Eq.~(\ref{est}) found by
\citet{KSW05,creminelli06,yadav08a,Smith06}. We consider two different
primordial shapes of non--Gaussianity: local non--Gaussianity, in which 
we expect the major 
contributions to the bispectrum  
from squeezed configurations, where $\ell_1\ll\ell_2,\ell_3$; and
equilateral non--Gaussianity, which predict large bispectrum 
contributions for equilateral shapes, $i.e.$ for $\ell_1\sim \ell_2\sim \ell_3$.
The corresponding $f_{\rm NL}$ estimators are described in the literature
cited above. In the following subsections we recall the definitions 
of the main quantities involved.

\subsubsection{local  estimator}
In the context of a local model of primordial non-Gaussianity it is useful 
to define the quantities 
\beqar
 \label{eq:alpha_l}
 \alpha_\ell(r)
 &\equiv&
  \frac{2}{\pi}\int k^2 dk g_{T\ell}(k) j_\ell(k r),\\
 \label{eq:beta_l}
  \beta_\ell(r) &\equiv& 
  \frac{2}{\pi}\int k^2 dk P(k) g_{T\ell}(k) j_\ell(k r),
\eeqar
where $g_{Tl}(r)$ is the radiation transfer function, $j_l(k r)$ is a Bessel 
function and $P(k)$ is the power--spectrum of the primordial perturbations.
With these quantities we can obtain two filtered maps from the 
harmonic coefficients $a_{\ell m}$ of the observed CMB:
\beqar
A(r,\hat{\mathbf {n}}) &\equiv&
  \sum_{\ell m} \frac{\alpha_\ell(r)W_\ell}{\tilde{C}_\ell}a_{\ell m}Y_{\ell m}(\hat{\mathbf {n}})\,,\label{mapA}\\
B(r,\hat {\mathbf {n}}) &\equiv&
  \sum_{\ell m} \frac{\beta_\ell(r)W_\ell}{\tilde{C}_\ell}a_{\ell m}Y_{\ell m}(\hat{\mathbf {n}})\,,\label{mapB}
\eeqar
where $W_\ell$ is the beam function of the instrument and $\tilde{C}_\ell$ is 
the theoretical power-spectrum $C_\ell$, after including effects of beam 
and noise as $\tilde{C}_\ell \equiv C_\ell W_\ell^2+\sigma_0^2$. Here we are 
approximating the noise as homogeneous $\left<n_{\ell m}n_{\ell'm'}^*\right>\simeq \sigma_0^2\delta_{\ell\ell'}\delta_{mm'}$.
With these maps we can construct a cubic statistics~\citep{KSW05}
\beq\label{eq:Sprim_loc}
{\mathcal S}^{\rm local}_{\rm prim} \equiv 4\pi \int r^2 dr 
  \int \frac{d^2\hat{\mathbf{n}}}{4\pi}
  A(r,\hat{\mathbf{n}}) B^2(r,\hat{\mathbf{n}}).
\eeq
This expression reduces exactly to Eq.~(\ref{eq:Sprim}), so we can define the
estimator of Eq.~(\ref{est}).
In the case of full sky coverage and homogeneous noise, the estimator is fully optimal,
$i.e.$ it saturates the Cramer-Rao bound, providing the minimum variance estimator.
But it is not realistic to consider a full sky coverage. If 
$f_{sky}$ is the fraction of sky observed we multiply the normalisation $N$ of
Eq.~(\ref{N_Sprim}) by $f_{sky}N$ \citep{KSW05,creminelli06}.
Moreover in an experiment the noise results anisotropic, since  different sky regions 
are observed for different amounts of time. Thus the noise covariance matrix 
$\left<n_{\ell m}n_{\ell'm'}^*\right>$ is no longer diagonal. The resulting correlation
between small and large scales causes
an increase of the variance of the estimator at high $\ell$'s.
This can be corrected with a linear term defined by 
\citep{creminelli06,yadav08a}
\beqar\label{Slin}
{\mathcal S}^{linear}_{\rm prim}&&=-\int r^2 dr \int d^2\hat{\mathbf{n}}
\Big\{2B(r,\hat{\mathbf{n}})
\langle A_{sim}(r,\hat{\mathbf{n}})\cdot\nonumber\\
 &&B_{sim}(r,\hat{\mathbf{n}})\rangle_{MC}
+\,A(r,\hat{\mathbf{n}})
\langle B^2_{sim}(r,\hat{\mathbf{n}})\rangle_{MC}\Big\},
\eeqar
where $A_{sim}(r,\hat{\mathbf{n}})$ and $B_{sim}(r,\hat{\mathbf{n}})$ are
the $A$ and $B$ maps generated from Monte Carlo simulations that contain
signal and noise, and $\langle\dots \rangle$ 
denotes average over the simulations. 
The estimator is finally given by
\beq\label{est:loc}
\hat{f}^{\rm local}_{\rm NL}= 
\frac{{\mathcal S}^{\rm local}_{\rm prim}+{\mathcal S}^{linear}_{\rm prim}}{N},
\eeq
This treatment greatly improves the final error bars of the estimator,
but it does not yet provide fully optimality. This is achieved by \citet{smith09}
through the implementation in Eq.~(\ref{est}) of the full covariance matrix.

\subsubsection{equilateral estimator}
In addition to $\alpha_\ell(r)$ and 
$\beta_\ell(r)$~(Eqs.~(\ref{eq:alpha_l},\ref{eq:beta_l})) we can define the functions
\begin{eqnarray}
\gamma_\ell(r) & \equiv & \frac2\pi \int k^2 dk {P(k)}^{1/3}g_{T\ell}(k) j_\ell(k r), \\
\delta_\ell(r) & \equiv & \frac2\pi \int k^2 dk {P(k)}^{2/3}g_{T\ell}(k) j_\ell(k r) \;.
\end{eqnarray}
and obtain the filtered maps $C(r,\hat{\mathbf{n}})$ and $D(r,\hat{\mathbf{n}})$ 
in the same way of the maps $A(r,\hat{\mathbf{n}})$ and $B(r,\hat{\mathbf{n}})$
(Eqs.~(\ref{mapA},\ref{mapB})).
Now the statistics is given by \citep{creminelli06}
\beqar
{\mathcal{S}}^{\rm equil}_{prim} &=& -18 \int  r^2 dr \int d^2 \hat{\mathbf {n}} \; \Big[ A(r, \hat{\mathbf {n}}) B^2(r, \hat{\mathbf {n}}) \nonumber\\
&+& \frac23 
D^3(r,\hat{\mathbf{n}})
- 2 B(r,\hat{\mathbf {n}})C(r,\hat{\mathbf {n}})
D(r, \hat{\mathbf {n}}) \Big] 
\eeqar
and the estimator results again
$\hat{f}^{\rm equil}_{\rm NL}= {\mathcal S}^{\rm equil}_{\rm prim}/N$,
where $N$ is given by Eq.~(\ref{N_Sprim}).
In this case the anisotropic noise is not an issue, since the spurious effect is 
quite low for equilateral bispectrum configurations.

\section{Analysis}\label{analysis}

In order to test the impact of the $1/f$ noise and the asymmetric beam  
we have analysed realistic {\sc Planck}
simulations of Gaussian CMB. The two systematic effects are studied separately,
considering two different sets of {\sc Planck}-like maps in which 
the $1/f$ noise or the asymmetric beam are simulated.
For each systematic the test 
consists in a comparison of the results obtained on the realistic
simulation set, with those obtained on another set of 
simulations hereafter dubbed ``nominal'', 
in which the considered systematic is not present.

\subsection{Simulation sets}
Realistic {\sc{Planck}} simulations can be obtained with the 
{\sc{LevelS}} pipeline \citep{reinecke06}, an assembly of numerical tools 
specifically developed to model the output data of the {\sc{Planck}} satellite.
{\sc{LevelS}} has been used to create
simulations of the time-ordered-data (TOD) for the 12 radiometers of the
LFI-70 GHz channel. The two systematic effects
considered are expected to be stronger for the LFI channels. 
The choice in particular of the 70 GHz channel in our study let us work with
the best sensitivity and angular resolution among the LFI channels.
The TODs cover 1 year of mission. 
Every set of TODs has been then converted in a map,
using the destriper map-making {\sc Springtide} \citep{ashdown07a}, with
$N_{side}=1024$\footnote{$i.e.$ $N_{pix}=12 N^2_{side}\sim10^7$, 
pixel size $\sim3.4'$} in the HEALPix pixelization scheme.
For each systematic effect we have a set of 100 {\sc Planck}-like maps.
The size of the set is limited because of the large resources
required by the TODs simulations and the subsequent map-making,  
both in terms of CPU computing time and disk space.
We have to keep in mind this limited size when looking at the results.
Nevertheless we have found that 100 maps are sufficient to obtain 
meaningful results.
We describe now the characteristics specific of each set.

\subsubsection{$1/f$ noise simulation set}
For each radiometer the instrumental noise is simulated as a sum of white 
and $1/f$
noise, with PSD given by Eq.~(\ref{PSDoof}). The parameters adopted are
reported in Table~\ref{oof:param}.
\begin{table}[!h]
\caption{parameters adopted in the $1/f$ noise simulations}\label{oof:param}
\centering
\begin{tabular}{rl}
\hline
standard deviation &$\sigma=1787.6\, \mu$K\\
knee frequency &$f_{k}=50$ mHz\\
$1/f$ slope &$\alpha=1.7$\\
sampling frequency &$f_s=76.8$ Hz\\
\hline
\end{tabular}
\end{table}
Due to the {\sc Planck} scanning strategy the noise is 
anisotropic. The white noise level chosen satisfies the goal sensitivity of 
the LFI-70
GHz channel \citep{bluebook}. Subsequent tests on the 70 GHz radiometers have shown a 
lower knee frequency. Moreover 
comparative map-making methods studies have found that the residual noise is slightly 
higher for {\sc{Springtide}} with respect to other destriping implementation
\citep{ashdown07a,ashdown07b,ashdown09}. Therefore 
these simulations are a conservative upper limit for the $1/f$ noise.

In addition to these 100 realistic noise maps, we have considered 
also a set of nominal simulations, $i.e.$ 100 white noise maps, 
in which the noise is a realization of 
anisotropic white noise with the same
nominal standard deviation
of the {\sc{Planck}} simulations.
To both noise map sets we have added 100 maps of Gaussian CMB sky.
The CMB maps are 
random realizations of the $WMAP$-1 year
best fit power spectrum, convolved  with the angular resolution of the LFI-70 GHz 
channel with a circular Gaussian beam of $14'$.
Finally, for the SMHW and the MF analyses we need Monte Carlo (MC) simulations
for calibration. Thus we have created 5000 MC maps of anisotropic white noise and
Gaussian CMB. We remark that the MC calibration does not contain the $1/f$ noise.

\subsubsection{asymmetric beam simulation set}
The {\sc{LevelS}} pipeline has been used to simulate a {\sc{Planck}}-like 
observation of 100 Gaussian CMB skies, created as 
random realizations of the $WMAP$-1 year best fit power spectrum.
The sky observation is simulated considering the best-fit elliptical beams of the
LFI-70 GHz, obtained in realistic main beam 
simulations \citep{sandri04}. The parameters of the bivariate Gaussian fitting 
for each radiometer are reported in Table~\ref{tab:asymbeams}. The twelve 
radiometers form symmetric couples due to the symmetry of the LFI focal plane.
\begin{table}[!h]
\caption{parameters adopted in the asymmetric beam simulations}\label{tab:asymbeams}
\centering
\begin{tabular}{l|cc}
\hline
id\footnote{LFI-70 GHz radiometers label; ``a'' and ``b'' refer to the polarization tuning.} & FWHM\footnote{Geometric mean of FWHM (full width at half maximum) of the major and minor axes of the beam ellipse (FWHM$=\sqrt{\rm fwhm_{max}fwhm_{min}}$).} & ellipticity\footnote{Ratio $\rm fwhm_{max}/fwhm_{min}$.}\\
\hline
18a, 23a& 13.0328&  1.2601\\
18b, 23b& 12.9916&  1.2184\\
19a, 22a& 12.7053&  1.2509\\
19b, 22b& 12.6494&  1.2162\\
20a, 21a& 12.4795&  1.2460\\
20b, 21b& 12.4273&  1.2224\\
\hline
\end{tabular}
\end{table}
The sampling in time has been also simulated. Since the effect is a 
widening of the beam in the scan direction, it has been reproduced using a
beam smeared in the rotation direction, perpendicular to the ecliptic plane.

\begin{figure}[!t]
\centering
\includegraphics[angle=90,width=7.5cm]{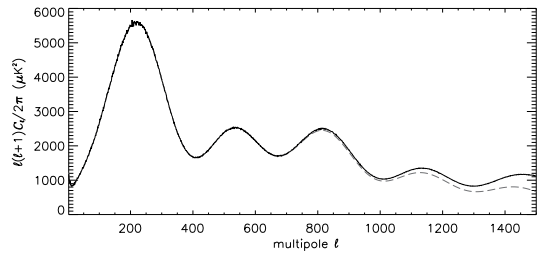}
\caption{Average power spectrum of 100 realistic simulations with asymmetric beam 
deconvolved with a circular Gaussian beam of $13'$ (black line). The theoretical 
CMB spectrum is shown with a dashed line.}\label{clasym}
\end{figure}

Furthermore we have created a set of 100 nominal simulations of Gaussian CMB, 
made again as random realizations of the $WMAP$-1 year best fit 
power spectrum, but in this case 
convolved with a circular Gaussian beam of $13'$. This is the 
resolution of the symmetric beam which best approximates 
the asymmetric one (Fig.~\ref{clasym}). 
For some analysis we have also added a random realization of anisotropic white noise
to each CMB map of both sets. The nominal standard deviation per sample is 
$\sigma=1787.6\, \mu$K (as in Table~\ref{oof:param}). We have finally created 
5000 MC simulations to calibrate the SMHW and the MF analyses,
following the same recipe of the nominal simulations.

\subsection{SMHW analysis}\label{smhw:analysis}

We have calculated the wavelets coefficients 
$w_i$ at 18 different scales $R$, 
starting from the pixel size:
$R=$ $3.4$, $6.9$, $10$, $13.7$, $25$, $50$, $75$, $100$, $150$, $200$, $250$,
 $300$, $400$, $500$, $600$, $750$, $900$, and $1050$ arcm.
For each scale we have computed standard deviation $\sigma(R)$, skewness $S(R)$ 
and kurtosis $K(R)$ of the distribution of the coefficients (Eqs.~(\ref{skewkurt})).
For both the $1/f$ noise and the asymmetric beam analysis, the statistics 
are calculated on the 100 realistic
simulations containing the systematic effect and on the 100 
nominal simulations without it. For calibration we have also calculated 
$\sigma(R)$, $S(R)$ and $K(R)$ over 5000 MC simulations.

Since the distribution of the MC values is close to Gaussian,
for every statistic and map we can construct a $\chi^2$ test in this form:
\beq\label{covchisq}
\chi^2(n)=\left({\bf x}(n)-\langle {\bf x}_{\rm MC}\rangle\right)^T {\bf C}^{-1}_{\rm MC}
\left({\bf x}(n)-\langle {\bf x}_{\rm MC}\rangle\right),
\eeq
where the vector ${\bf x}$ has elements $x_i=\sigma(R_i)$, $S(R_i)$ or $K(R_i)$ 
and ${\bf C}^{-1}_{\rm MC}$ is the MC covariance matrix $C_{ij}=\langle
\left(x_i-\langle x_i\rangle\right)\left(x_j-\langle x_j\rangle\right)\rangle$.
The $\chi^2$ test is performed on each map of the realistic and of the nominal
simulation sets, with $n=1-100$, and on each MC map, with $n=1-5000$.
Moreover we have established the acceptance intervals at different significance 
levels $\alpha$ for each statistic $\sigma(R)$, $S(R)$ or $K(R)$
and scale. Fixing $R$, we look
at the distribution of the values over the 5000 MC maps: 
an acceptance interval 
is defined as the one containing a percentage $1-\alpha$ of values, where the
remaining percentage is the same above and below the interval, $i.e.$ $\alpha/2$
at each side. We have chosen three significance levels: $68$\% ($1\sigma$), 
$5$\% ($2\sigma$) and $1$\%. When a value falls out of an interval,
the significance of the non-Gaussianity indication is given by 
$\alpha$: lower is the significance level $\alpha$, stronger is the indication.

With these quantities we have constructed three figures of merit.
Each figure is evaluated for standard deviation $\sigma(R)$, skewness $S(R)$ 
and kurtosis $K(R)$ on both the realistic and the nominal simulation sets.
We have looked at all the three statistics, but remembering that only skewness
and kurtosis are Gaussian estimators. We have also studied  
the standard deviation for
a better characterisation of the systematic effect.
\begin{enumerate}
\item $\chi^2$ distributions comparison:\\
We have tested if the 100 $\chi^2$ values of the simulation set 
belong to the same
distribution of 5000 $\chi^2_{\rm MC}$ values of the MC simulations. 
The degree of agreement is quantified as the percentage of  
$\chi^2_{\rm MC}$ greater than the mean of the $\chi^2$  of the 100 simulations;
\item number of $\chi^2$ out of $2\sigma$:\\
For each $\chi^2$ value of the simulation set we have looked at the 
percentage of MC simulations with a lower 
$\chi^2_{\rm MC}$. If this percentage is less than $2.5$\% or more than $97.5$\%, 
then the $\chi^2$ value lies outside  $2\sigma$ of the $\chi^2_{\rm MC}$ 
distribution;
\item number of outliers:\\
For each scale we have counted the number of maps with statistic of the wavelet
coefficients out of the acceptance intervals at $5$\% and $1$\%.
\end{enumerate}
The impact of the $1/f$ noise or the asymmetric beam has been tested comparing
the figures of merit of the realistic simulation set against 
the figures of the nominal one.

\subsubsection{SMHW: $1/f$ noise results}

\begin{figure}[!h]
\centering
\includegraphics[angle=90,width=0.5\textwidth]{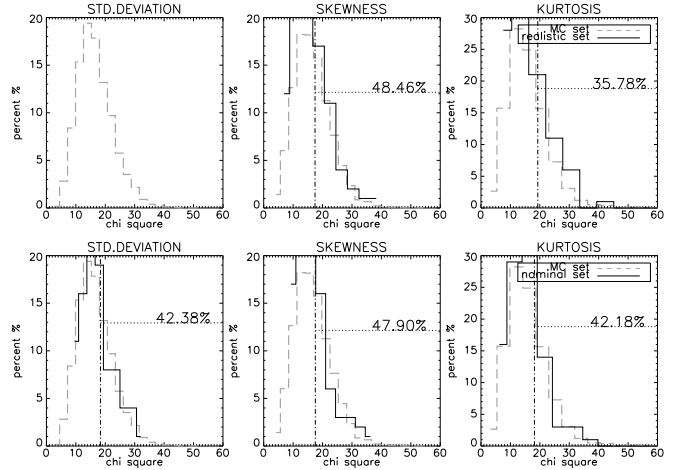}
\caption{SMHW - $1/f$ noise results: $\chi^2$ distributions comparison
for realistic (top line) and nominal (bottom line) simulations.
The dot-dashed line shows
the $\langle \chi^2 \rangle$ of the simulation set. We indicate 
the percentage of $\chi^2_{\rm MC}$ greater than this mean.}
\label{case1figmerI}
\end{figure}

The first figure of merit for the $1/f$ noise is showed in Fig.~\ref{case1figmerI}.
The impact of the $1/f$ is noticeable only in the standard deviation: the
$\chi^2$ distribution of the realistic set is not displayed in the plot 
because the values are out of range respect to the MC set 
 (upper-left plot in Fig.~\ref{case1figmerI}).
This is indeed expected, because of the increase of the noise variance with respect 
to the white noise level of the MC calibration due to the 
 $1/f$ noise contribution.
All the other plots show an agreement between the set of maps and the 
MC calibration, indicated also by the percentage of $\chi^2_{\rm MC}$ greater than the  $\chi^2$ mean. Table~\ref{case1figmerII} reports the second figure of merit.
 Apart again from the standard deviation, the $1/f$ noise does not increase
the occurrence of skewness or kurtosis out of $2\sigma$ from the Gaussian
behavior as calibrated with MC simulations.

\begin{table}[!h]
\caption{SMHW - $1/f$ noise results: number of $\chi^2$ out of $2\sigma$}\label{case1figmerII}
\centering
\begin{tabular}{c|ccc}
\# of $\chi^2$ & std.deviation & skewness & kurtosis\\
\hline
realistic &100&3&3\\
nominal &1&5&7\\
\end{tabular}
\end{table}

\begin{figure}[!h]
\centering
\includegraphics[angle=90,width=0.5\textwidth]{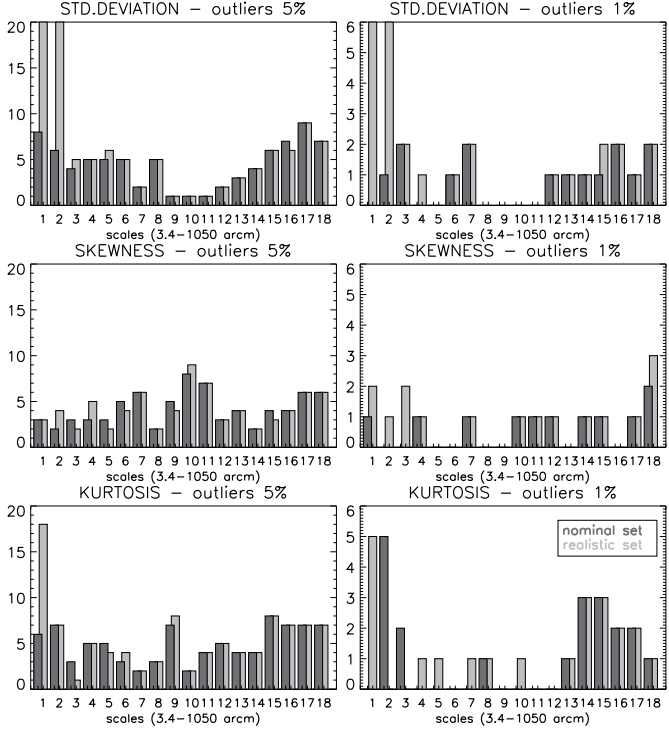}
\caption{SMHW - $1/f$ noise results: number of outliers of the $5$\% and $1$\% acceptance intervals.}
\label{case1figmerIII}
\end{figure}

The third figure of merit  (Fig.~\ref{case1figmerIII}) shows the impact of the 
$1/f$ noise on the different scales. We can observe that at 
scales $R \sim$ pixel scale
the presence of $1/f$ noise increase the probability for a map 
to have the standard deviation of wavelet coefficients out of $1$\% or $5$\%.
Actually this scale is noise dominated and usually it is not considered in Gaussianity analysis.

Before moving on in the analysis, a general comment on 
the third figure of merit is mandatory. We can sometime observe a number 
of outliers greater than what expected in particular for the nominal set, since 
these maps are obtained with the same pipeline of the MC simulations. 
This is only due to the limited size of the simulation sets, which increases the 
number of values in the distribution tails.

We have repeated the analysis also with an amplified noise 
contribution: in both the realistic and the nominal simulation sets the noise maps
have been multiplied by a factor equal to $2.5$. The results found are qualitatively unchanged,
thus we do not show them here.

\begin{figure}[!h]
\centering
\includegraphics[angle=90,width=0.5\textwidth]{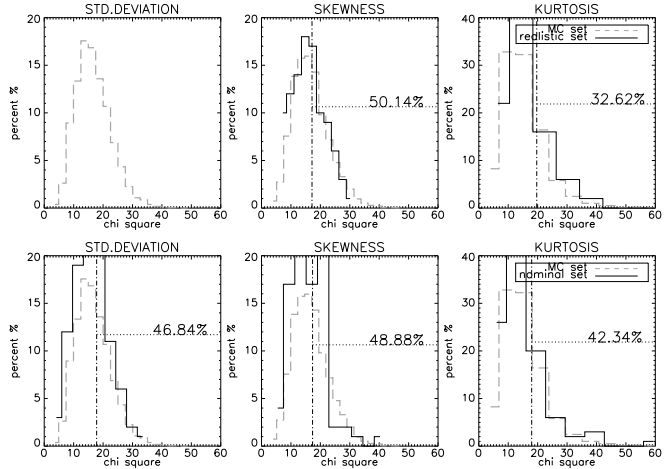}
\caption{SMHW - $1/f$ noise results: $\chi^2$ distributions comparison [only noise simulations] for realistic (top line) and nominal (bottom line) simulations.
The dot-dashed line shows
the $\langle \chi^2 \rangle$ of the simulation set. We indicate 
the percentage of $\chi^2_{\rm MC}$ greater than this mean.}
\label{case3figmerI}
\end{figure}

\begin{figure}[!h]
\centering
\includegraphics[angle=90,width=0.5\textwidth]{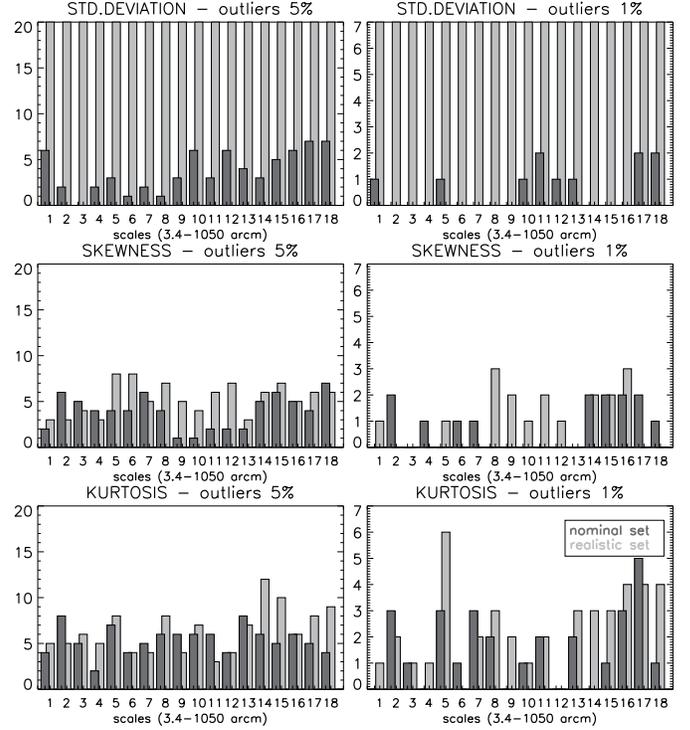}
\caption{SMHW - $1/f$ noise results: number of outliers [only noise simulations]}
\label{case3figmerIII}
\end{figure}

For a better understanding of the noise properties we finally have analysed only 
noise simulation, in which no CMB signal is added. We can look at the $\chi^2$ test
of the realistic simulations
in Fig.~\ref{case3figmerI}. Apart from the standard deviation, where again we have 
obtained values out of MC set range, 
only the kurtosis shows a somewhat lower percentage for the realistic set 
respect to the nominal one. Anyway the value still indicates a good agreement
with the MC distribution.  Also the number
of $\chi^2$ out of $2\sigma$ reported in Table~\ref{case3figmerII}
is not affected by the $1/f$ noise contribution.
But when we look at the third figure of merit in Fig.~\ref{case3figmerIII}, 
we can see that the $1/f$ noise
increases the number of outliers. The standard deviation is affected at all the scales,
since here the increase of noise variance is no more hidden by any CMB signal.
Also skewness and kurtosis are affected at some scales, but we observe that 
 the number of outliers remains within statistically reasonable limits.

\begin{table}[!h]
\caption{SMHW - $1/f$ noise results: number of $\chi^2$ out of $2\sigma$
[only noise simulations]}\label{case3figmerII}
\centering
\begin{tabular}{c|ccc}
\# of $\chi^2$ & std.deviation & skewness & kurtosis\\
\hline
realistic&100&3&4\\
nominal&4&4&7\\
\end{tabular}
\end{table}

\subsubsection{SMHW: asymmetric beam results}

\begin{figure}[!h]
\centering
\includegraphics[angle=90,width=0.5\textwidth]{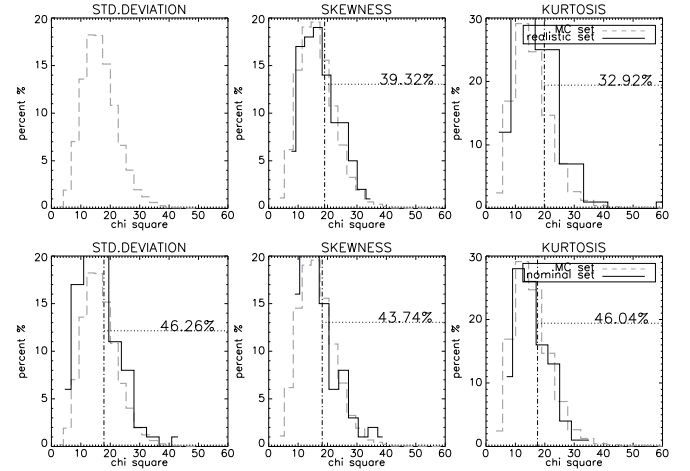}
\caption{SMHW - asymmetric beam results: $\chi^2$ distributions comparison
for realistic (top line) and nominal (bottom line) simulations.
The dot-dashed line shows
the $\langle \chi^2 \rangle$ of the simulation set. We indicate 
the percentage of $\chi^2_{\rm MC}$ greater than this mean.}
\label{abcase1figmerI}
\end{figure}
The first figure of merit in Fig.~\ref{abcase1figmerI} shows that the asymmetric
beam has an impact on the standard deviation: the
$\chi^2$ values of the realistic set are out of range with respect to the MC set values.
We can see in Fig.~\ref{clasym} that 
indeed the elliptical beam used in the realistic simulation produces a different 
smoothing with respect to the circular beam, used in the nominal and in the 
MC simulations. Instead skewness and kurtosis percentages indicate a still
good agreement with the MC calibration, even if lower with respect to the 
symmetric beam maps. The number of $\chi^2$ out of $2\sigma$ in 
Table~\ref{abcase1figmerII} confirms that
the asymmetric beam has a significant 
impact on the $\chi^2$ distribution of the standard deviation statistic only.
\begin{table}[!h]
\caption{SMHW - asymmetric beam results: number of $\chi^2$ out of $2\sigma$}\label{abcase1figmerII}
\centering
\begin{tabular}{c|ccc}
\# of $\chi^2$ & std.deviation & skewness & kurtosis\\
\hline
realistic&100&3&6\\
nominal&8&4&4\\
\end{tabular}
\end{table}
\begin{figure}[!h]
\centering
\includegraphics[angle=90,width=0.5\textwidth]{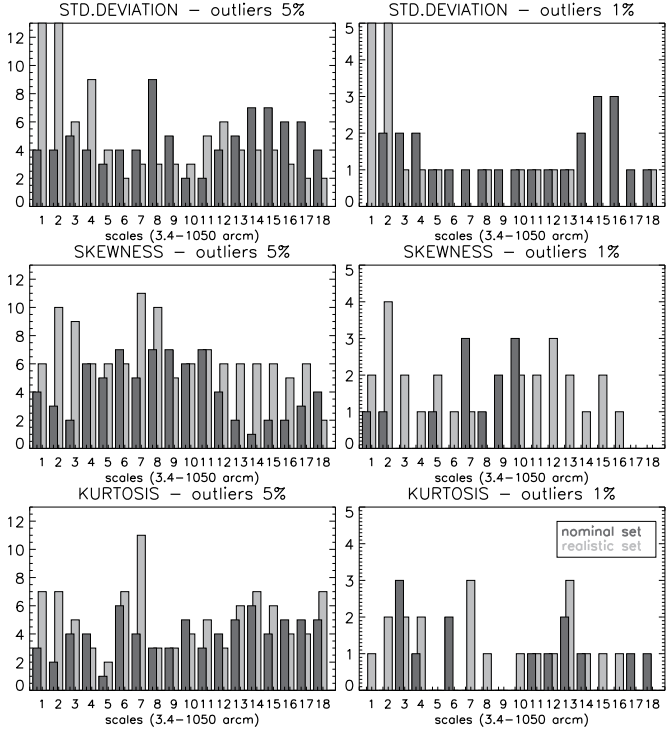}
\caption{SMHW - asymmetric beam results: number of outliers}
\label{abcase1figmerIII}
\end{figure}
Looking at the number of outliers in Fig.~\ref{abcase1figmerIII} we observe
that the asymmetric beam increases the occurrence of wavelet
coefficients statistics out of the acceptance intervals. This happens only at small
scales for the standard deviation, while we have a general slight increase 
for kurtosis and skewness. Anyway for these two statistics the numbers of outliers
are still admissible and the departure from the MC Gaussian 
calibration is not significant. 
We have repeated the analysis adding an anisotropic white noise to the CMB maps,
both in the realistic and in the nominal simulation sets. We have found that 
the noise does
not modify the impact of the asymmetric beam and the results do not change.

\subsection{MF  analysis}

In order to investigate different angular scales, for both the $1/f$ noise
and the asymmetric beam study we have smoothed the sets of 100
simulations, realistic and nominal, with Gaussian beams at $0$ ($i.e.$ 
no smoothing), $10$, $30$, $60$, $120$, $180$ and $240$ arcm. 
For every map and smoothing scale we then have
estimated the MF, $i.e.$ area, length, genus and skeleton length. 
The derivative computation is performed up to
an appropriate multipole $\ell_{max}$ for each scale: from $850$ to $60$ 
for the larger beam. The MF are estimated for $200$ threshold values 
$\nu=-4\sigma\div4\sigma$, but in the subsequent analysis we consider
only thresholds  $-3\sigma<\nu<3\sigma$ ($-2.5\sigma<\nu<2.5\sigma$
for the two larger scales). We have applied the same procedure to the 5000
MC maps for calibration. Having verified the Gaussianity of the values
distribution over the MC, we have constructed a $\chi^2$ test. We 
have performed a diagonal
$\chi^2$ test, due to the narrow step-size
used in the thresholds range, defining
\beq\label{diagchi}
\chi^2(n)=\sum_\nu\left[\frac{f(\nu,n)-\langle f_{\rm MC}(\nu) \rangle}{\sigma_{\rm MC}(\nu)}\right]^2,
\eeq
where $f$ is one of the MF, $\langle f_{\rm MC}(\nu) \rangle$ and 
$\sigma_{\rm MC}(\nu)$ are the mean and the standard deviation over the MC simulations. The $\chi^2$ test has been performed on the realistic and the nominal simulation
sets and on the MC calibration maps.

The evaluation of the results is achieved analysing two figures of merit,
identical to the first two figures considered in the SMHW analysis in 
\S\ref{smhw:analysis}, $i.e.$:
\begin{enumerate}
\item $\chi^2$ distributions comparison;
\item number of $\chi^2$ out of $2\sigma$.
\end{enumerate}
The figures of merit are evaluated for each smoothing scale and functional. 
For both the $1/f$ noise and the asymmetric beam study, we are going to 
compare the results of the figures obtained on the realistic and on the nominal 
simulation set, $i.e.$ respectively with or without the systematic considered.

\subsubsection{MF: $1/f$ noise results}

\begin{figure}[!h]
\centering
\includegraphics[angle=90,width=0.5\textwidth]{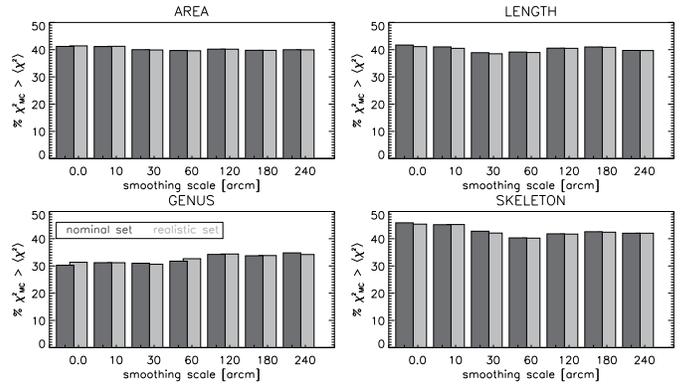}
\caption{MF - $1/f$ noise results: $\chi^2$ distributions comparison}
\label{minko1fig1}
\end{figure}
\begin{figure}[!h]
\centering
\includegraphics[angle=90,width=0.5\textwidth]{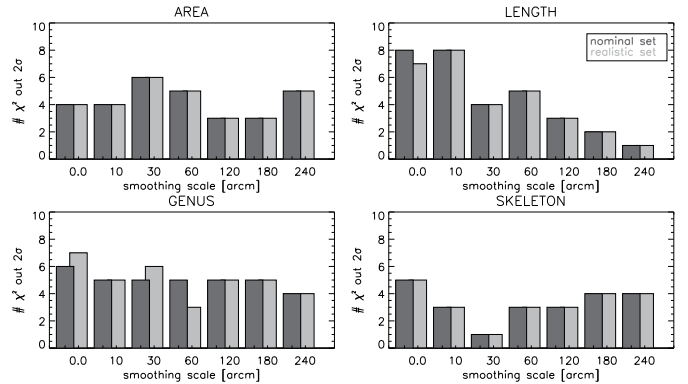}
\caption{MF - $1/f$ noise results: number of $\chi^2$ out of $2\sigma$.}
\label{minko1fig2}
\end{figure}
In the first figure of merit (Fig.~\ref{minko1fig1}) it is straightforward to see
that there is not any noticeable difference between the results obtained with 
the $1/f$ noise included and the 
results from the nominal simulations, in
none of the smoothing scales applied to the maps.
Also looking at the number of $\chi^2$ out of 
$2\sigma$ in Fig.~\ref{minko1fig2}, 
we can say that the agreement of the simulation set with the MC calibration
is not affected by the $1/f$ noise. We have checked the robustness of  
the results repeating the analysis on simulations with an higher noise level.
The noise maps, both in the nominal and in the realistic simulation set, have 
been multiplied by a factor of $2.5$. The results found
are almost identical to that showed.

\subsubsection{MF: asymmetric beam results}

\begin{figure}[!h]
\centering
\includegraphics[angle=90,width=0.5\textwidth]{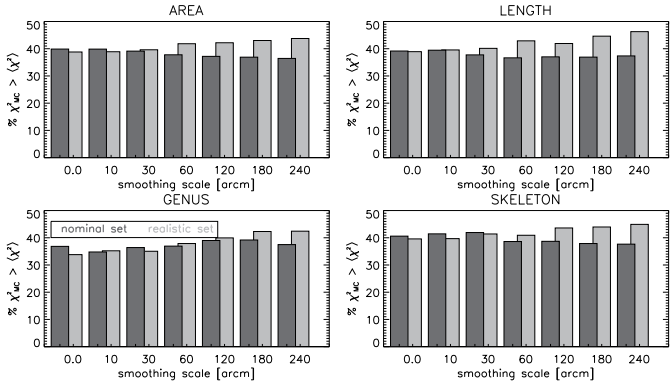}
\caption{MF - asymmetric beam results: $\chi^2$ distributions comparison}
\label{abminko1fig1}
\end{figure}

\begin{figure}[!h]
\centering
\includegraphics[angle=90,width=0.5\textwidth]{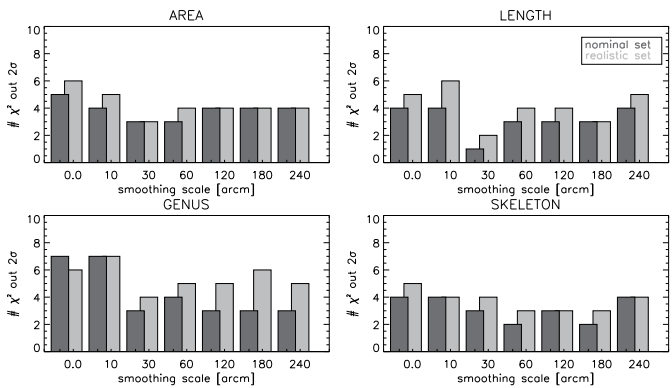}
\caption{MF - asymmetric beam results: number of $\chi^2$ out of $2\sigma$}
\label{abminko1fig2}
\end{figure}

Comparing the distribution of $\chi^2$ values of the simulation sets
with the MC calibrations (Fig.~\ref{abminko1fig1}), we observe that the
asymmetric beam of the realistic simulations does not cause a lower agreement
with respect to the nominal simulations. In Fig.~\ref{abminko1fig2} we can see
that for the asymmetric beam maps in many cases we have found more  
$\chi^2$ values out of $2\sigma$ respect to the symmetric beam maps,
but the increase is not remarkable. We have also analysed the same 
CMB simulations after adding anisotropic white noise, but the results 
do not change significantly.

\subsection{$f_{\rm NL}$ estimator analysis}
\begin{figure}[!h]
\centering
\includegraphics[angle=90,width=0.5\textwidth]{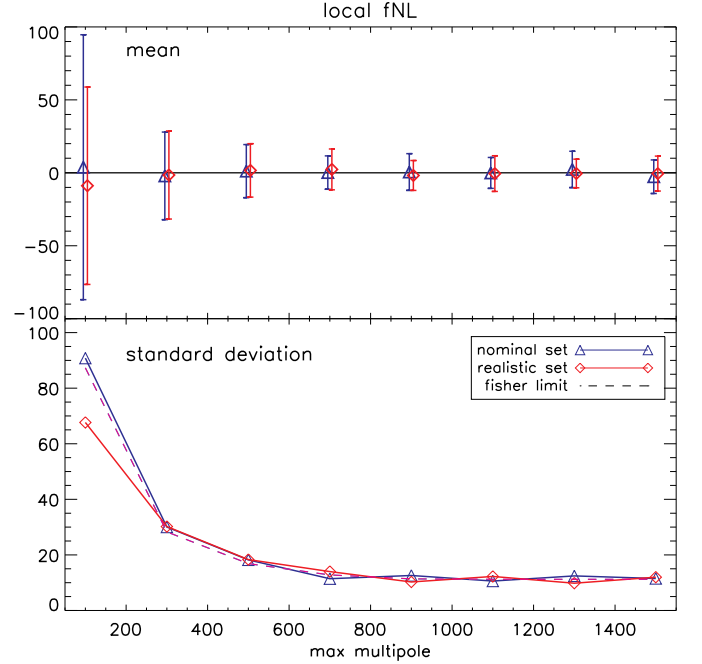}
\caption{$f_{\rm NL}^{\rm local}$ - $1/f$ noise results. Error bars and standard deviations are $1\sigma$.}\label{fnloc:oof}
\end{figure} 
\begin{figure}[!h]
\centering
\includegraphics[angle=90,width=0.5\textwidth]{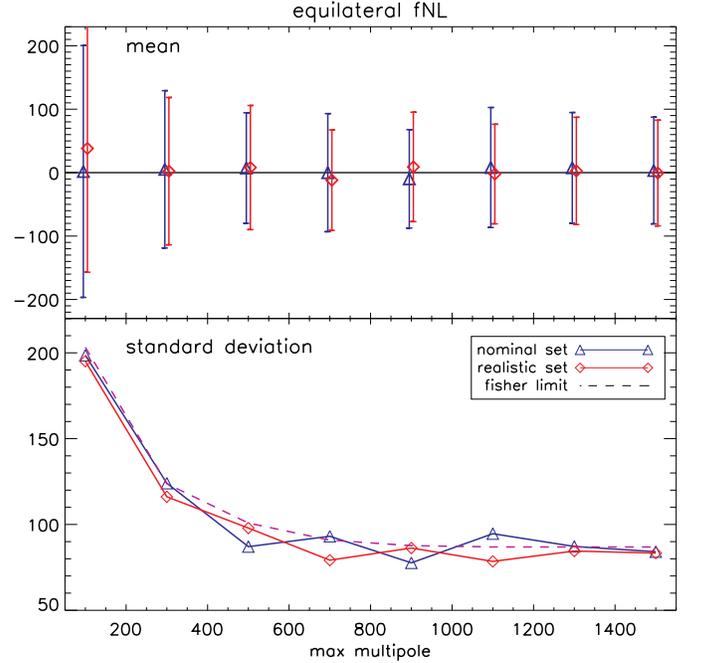}
\caption{$f_{\rm NL}^{\rm equil}$ - $1/f$ noise results}\label{fnleq:oof}
\end{figure} 
\begin{figure}[!h]
\centering
\includegraphics[angle=90,width=0.5\textwidth]{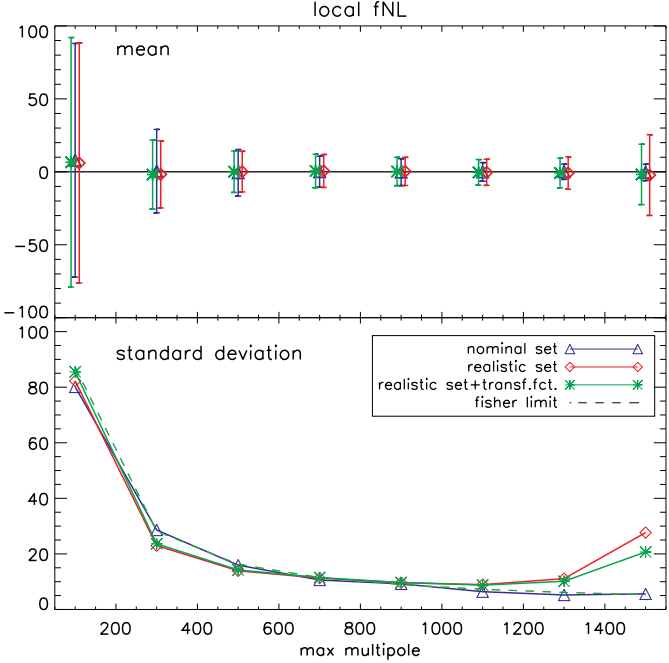}
\caption{$f_{\rm NL}^{\rm local}$ - asymmetric beam results}\label{fnloc:asbm}
\end{figure} 
\begin{figure}[!h]
\centering
\includegraphics[angle=90,width=0.5\textwidth]{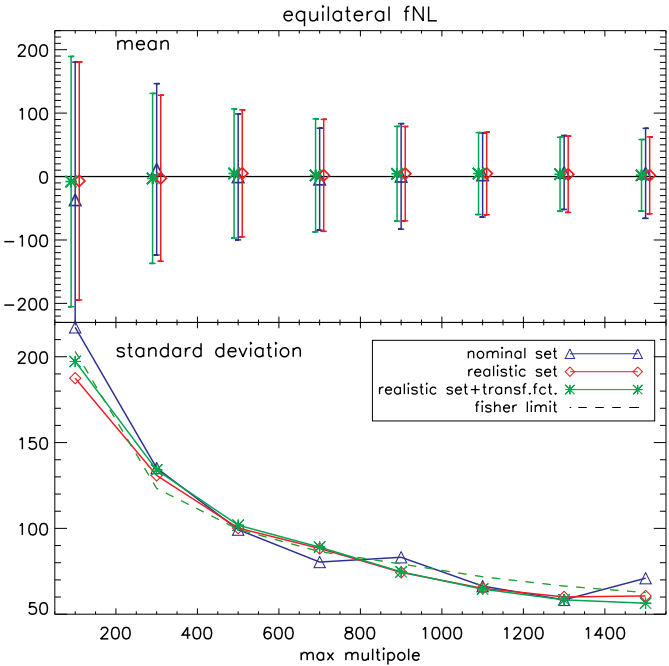}
\caption{$f_{\rm NL}^{\rm equil}$ - asymmetric beam results}\label{fnleq:asbm}
\end{figure} 
\begin{figure}[!h]
\centering
\includegraphics[angle=90,width=0.5\textwidth]{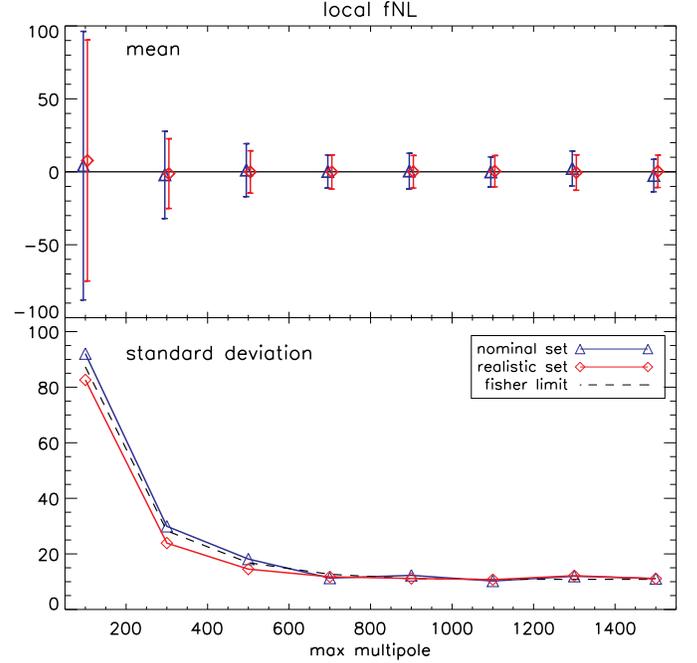}
\caption{$f_{\rm NL}^{\rm local}$ - asymmetric beam results [with anisotropic noise]}\label{fnloc:asbm_nn}
\end{figure} 

In order to test the impact of the $1/f$ noise and the asymmetric beam 
on the $f_{\rm NL}$ parameter, we have performed the estimation with the
cubic statistic described in \S\ref{fnlestimator}
 on the realistic and the nominal simulation sets, in both the local and 
the equilateral configuration. 
In the local configuration, when the noise is added, we have also corrected
for correlations introduced by 
the anisotropic noise, computing the linear term of
Eq.~(\ref{Slin}).
The bispectrum is calculated up to different
maximum values of multipole $\ell_{max}$. 
We have proceeded looking at the mean of the $f_{\rm NL}$ values obtained,
to check against biases with respect to the expected null value.
Then we have investigated the impact of the systematic
effects on the variance of the estimator, comparing the standard 
deviations of the two
sets of  100 $f_{\rm NL}$ values obtained from
the realistic and the nominal simulations.

\subsubsection{$f_{\rm NL}$: $1/f$ noise results}

As explained in \S\ref{fnlestimator}, correlation properties of the 
noise random field have to be properly accounted for, introducing correction
terms, like the linear term of Eq.~(\ref{Slin}); or
computing the full noise covariance matrix and 
performing an inverse
covariance matrix weighting of the data \citep{smith09}.
But in the case of the $1/f$ noise the estimation of the noise covariance matrix
is numerically challenging, in particular at the high resolution reached with
{\sc{Planck}} (see \citet{keskitalo09} for the study of the noise covariance matrix
for {\sc{Planck}} low-resolution data analysis). 
Before making any attempt in this direction, it is therefore useful to evaluate
the effective impact of the $1/f$ noise without accounting for it 
in the $f_{\rm NL}$ estimation. 
Actually the noise is considered in the computation of the Fisher matrix $N$ 
(Eq.~\ref{N_Sprim}) and of the maps $A(r,\hat{\mathbf{n}})$, $B(r,\hat{\mathbf{n}})$, $C(r,\hat{\mathbf{n}})$ and $D(r,\hat{\mathbf{n}})$ 
(Eqs.~(\ref{mapA},~\ref{mapB})), but it is accounted by means of its power 
spectrum, therefore
without considering the correlations introduced by the $1/f$ noise in the 
realistic simulations.

In Fig.~\ref{fnloc:oof} we show the results for the local configuration, while
in Fig.~\ref{fnleq:oof} there are the equilateral configuration results.
In both the cases we can not observe any bias in the mean of $f_{\rm NL}$. 
This is indeed expected, since the $1/f$ noise is correlated but still Gaussian.
But it is important to check against possible biases introduced by the not proper 
noise treatment itself.
Moreover the $1/f$ noise does
not increase the $f_{\rm NL}$ standard deviation, and the error bars have the 
same amplitude obtained from the nominal simulations.
The small deviations between the obtained values and the limit given by the Fisher matrix 
are compatible due to the limited size of the simulation sets.

\subsubsection{$f_{\rm NL}$: asymmetric beam results}

We can repeat here the same considerations made for the $1/f$ noise. The asymmetric
beam can introduce correlations in the data, but the numerical description of
this effect is demanding. Therefore we want to evaluate the effective impact of
this systematic  when it is not properly accounted for.
In the estimation of $f_{\rm NL}$ the beam considered in the computation 
is circular. In particular we have considered a circular Gaussian beam of $13'$, 
the same used in the nominal simulation. 

The results obtained for the
local and the equilateral $f_{\rm NL}$ are shown in Fig.~\ref{fnloc:asbm}
and Fig.~\ref{fnleq:asbm} respectively. Looking at the mean  of the $f_{\rm NL}$ 
values, not any bias is observable with respect to the null value 
in both the configurations.
For the equilateral case, the standard deviation is also not affected by the
asymmetric beam. In contrast in the local configuration 
we can observe an increase of the standard deviation at high multipoles.

In order to understand if this increase is due only to the different smoothing 
of the elliptical beam with respect to the circular one, we have
applied a transfer function in the $f_{\rm NL}$ estimation. 
This function is calculated as the mean of the spectra of the
realistic simulation set, divided by the theoretical spectrum 
convolved with the circular $13'$ beam. 
Therefore the ``effective'' beam, obtained as the circular beam multiplied 
for this transfer function, 
has the same amplitude of the elliptical beam. 
We have used this ``effective'' beam 
in the $f_{\rm NL}$ estimation of the asymmetric beam maps. 
The results obtained using the 
transfer function are shown in green color in the same Figs. \ref{fnloc:asbm}
and \ref{fnleq:asbm}.
We can observe that, even if lower, the increase in the standard deviation
of the local configuration is still present.

Then we have analyzed the same simulations after
adding an anisotropic white noise contribution to the CMB maps. 
This is indeed a more realistic situation. The noise does not change the 
results for the equilateral configuration. On the other hand the standard
deviations of the local  $f_{\rm NL}$ are significantly 
improved with respect to the situation without noise. 
We show that the local configuration results in Fig.~\ref{fnloc:asbm_nn}:
the increase of the error bars at high multipoles is no longer
observable.

\section{Conclusions}\label{conclus}
We have studied the impact of  the
$1/f$ noise and the asymmetric beam on searches for non-Gaussian primordial signals.
These two systematics effects are indeed expected to be present in the forthcoming 
{\sc{Planck}} CMB data. With realistic {\sc{Planck}}-like simulations we have evaluated
if they can affect the non-Gaussianity analysis carried out with 
different statistical approach:
blind tests,
as the SMHW and the Minkowski functionals, and an $f_{\rm NL}$ estimator for both
local and equilateral configurations, 
known as the ``KSW'' estimator. 
The two systematics have been studied separately.
In the case of the blind tests we have looked at different 
figures of merit, checking for false non-Gaussian detections,
while for the $f_{\rm NL}$ estimator we have checked for the bias
in the estimated value and for the increase in the variance.
In each case the results on realistic simulations have been compared with the results on
simulations without the systematic effect examined.

For both the $1/f$ noise and the asymmetric beam analysis, no tests conducted 
shows a significant impact on the final results. 
The $1/f$ noise can slightly affect only the SMHW analysis, and only when the noise
dominates over the CMB signal, $e.g.$ at pixel scale.
The blind tests are not significantly affected by the non circular beam,
but in the case of the $f_{\rm NL}$ estimator we can observe 
an increase of the variance of the local  $f_{\rm NL}$  at high multipoles,
when we consider simulations without noise.
Anyway this effect is canceled once we consider more realistic noisy simulations.

In conclusion, assuming that the simulations 
analysed well reproduce the actual
{\sc{Planck}} observations, if any non--Gaussian 
signal will be detected in the forthcoming {\sc{Planck}} CMB data with these 
statistical tests, it will not be due to the 1/f noise or the asymmetry of the beam.

\acknowledgements
We acknowledge Mark Ashdown and Torsti Poutanen for the 
production of the realistic simulation sets. We thank Hans Kristian Eriksen
for his help in developing our MF algorithm. FKH is grateful for an OYI grant from the Research Council of Norway.
The results described in this paper have been produced
using the Titan High Performance Computing facilities at the University of Oslo
({\tt{http://hpc.uio.no/index.php/RCS}}).


\begin{thebibliography}{}
\bibitem[Acquaviva et al.(2003)]{Acquaviva03} 
Acquaviva V., Bartolo N., Matarrese S., and Riotto A., Nucl.\ Phys.\ B 667 (2003) 119.

\bibitem[Ashdown et al.(2007a)]{ashdown07a}
Ashdown M.A.J. et al., \aap\ 467 (2007) 761.

\bibitem[Ashdown et al.(2007b)]{ashdown07b} 
Ashdown M.A.J. et al., \aap\ 471 (2007) 361.

\bibitem[Ashdown et al.(2009)]{ashdown09} 
Ashdown M.A.J. et al., \aap\ 493 (2009) 753.

\bibitem[Babich et al.(2004)]{babich04} 
Babich D., Creminelli P. and Zaldarriaga M., JCAP 8 (2004) 9.

\bibitem[Bartolo et al.(2001)]{BartoloReview} 
Bartolo N., Komatsu E., Matarrese S., and Riotto A., \physrep\ 402 (2004) 103.

\bibitem[Brossard et al.(2004)]{brossard04} 
Brossard J. et al., Proc.\ of the 5th International Conference on Space Optics (ICSO 2004), ESA SP-554 (2004) 333.

\bibitem[Cay\'on et al.(2000)]{Cayon00} 
Cay\'on et al., \mnras\ 315 (2000) 757.

\bibitem[Chen et al.(2007a)]{chen07a}
Chen X., Huang M.-X., Kachru S., and Shiu G., JCAP 0701 (2007) 002. 

\bibitem[Chen et al.(2007b)]{chen07b}
Chen X., Easther R., and Lim E.A., JCAP 0706 (2007) 023. 

\bibitem[Creminelli et al.(2006)]{creminelli06} 
Creminelli P., Nicolis A., Senatore L., Tegmark M., and Zaldarriaga M., JCAP 0605 (2006) 4. 

\bibitem[Curto et al.(2009)]{curto08} 
Curto, A., Mart\'inez-Gonz\'alez E., Mukherjee P., Barreiro R.B., Hansen F.K., Liguori M., and Matarrese S., \mnras\ 393 (2009) 615.

\bibitem[Efstathiou(2005)]{Efstathiou05} 
Efstathiou G., \mnras\ 356 (2005) 1549.

\bibitem[Eriksen et al.(2004)]{eriksen04} 
Eriksen H.K., Novikov D.I., Lilje P.B., Banday A.J., and G\'orski K.M., \mnras\ 612 (2004) 64.

\bibitem[G\'orski et al.(2005)]{healpix} 
G\'orski K.M., Hivon E., Banday A.J., Wandelt B.D., Hansen F.K., Reinecke M., and Bartelmann M., \apj\ 622 (2005) 759.

\bibitem[Hansen et al.(2009)]{Hansen08} 
Hansen F.K., Banday A.J., G\'orski K.M., Eriksen H.K., and Lilje P.B., \apj\ 704 (2009) 1448 (arXiv:0812.3795). 

\bibitem[Holman \& Tolley(2008)]{holmtol08}
Holman R. and Tolley A.j., JCAP 0805 (2008) 001. 

\bibitem[Janssen et al.(1996)]{janssen96} 
Janssen M. et al., Report PSI-96-01, FIRE-96-01 (1996) (arXiv:astro-ph/9602009).

\bibitem[Keih\"anen et al.(2004)]{elina04}
Keih\"anen E., Kurki-Suonio H., Poutanen T., Maino D., and Burigana C., \aap\ 428 (2004) 287.

\bibitem[Keskitalo et al.(2009)]{keskitalo09}
Keskitalo R. at al., (2009), arXiv:0906.0175.

\bibitem[Komatsu \& Spergel(2001)]{KomSperg01} 
Komatsu E. and Spergel D.N., \prd\ 63 (2001) 063002.

\bibitem[Komatsu et al.(2005)]{KSW05} 
Komatsu E., Spergel and Wandelt B.D., \apj\ 634 (2005) 14. 

\bibitem[Komatsu et al.(2009)]{komatsu08} 
Komatsu E. et al., \apjs\ 180 (2009) 330.

\bibitem[Langlois et al.(2008)]{langl08}
Langlois D., Renaux-Petel S., Steer D.A., and Tanaka T., \prd\ 78 (2008) 063523.

\bibitem[Linde \& Mukhanov(1997)]{lindemu97}
Linde A. and Mukhanov V., \prd\ 56 (1997) 535. 

\bibitem[Lyth et al.(2003)]{lyth03}
Lyth D.H., Ungarelli C, and Wands D., \prd\ 67 (2003) 023503.  

\bibitem[Maino et al.(2002)]{maino02}
Maino D., Burigana C., G\'orski K. M., Mandolesi N., and  Bersanelli M., \aap\ 387 (2002) 356.

\bibitem[Maldacena(2003)]{Maldacena03}
Maldacena J., JHEP 0305 (2003) 013.

\bibitem[Mart\'inez-Gonz\'alez et al.(2002)]{martinez02} 
Mart\'inez-Gonz\'alez E., Gallegos J.E., Arg\"ueso F., 
Cay\'on L., and Sanz J.L, \mnras\ 336 (2002) 22.

\bibitem[McEwen et al.(2008)]{mcewen08} 
McEwen J.D., Hobson M.P., Lasenby A.N., and Mortlock D.J., \mnras\ 388 (2008) 659.

\bibitem[Minkowski(1903)]{minkowski1903} 
Minkowski H., Mathematische Annalen 57 (1903) 447.

\bibitem[Pietrobon et al.(2009)]{pietrobon08} 
Pietrobon D., Cabella P., Balbi A., de Gasperis G., and Vittorio N., \mnras\ 396 (2009) 1682.
 
\bibitem[Reinecke et al.(2006)]{reinecke06} 
Reinecke M., Dolag K., Hell R., Bartelmann M., and En\ss lin T.A., \aap\ 445 (2006) 373.

\bibitem[Rudjord et al.(2009)]{rudjord09} 
Rudjord \O, Hansen F.K., Lan X., Liguori, M., Marinucci D., and Matarrese S., \apj\ 701 (2009) 369 (arXiv:0901.3154).

\bibitem[Sandri et al.(2004)]{sandri04} 
Sandri M., Villa F., Mandolesi N., Bersanelli M., and Nesti R., Optical, Infrared, and Millimeter Space Telescopes., \procspie\ 5487 (2004) 532.

\bibitem[Senatore et al.(2009)]{senatore09}
Senatore L., Smith K.M., and Zaldarriaga M., (2009), arXiv:0905.3746.

\bibitem[Smith \& Zaldarriaga(2006)]{Smith06}
Smith K.M. and Zaldarriaga M., arXiv:astro-ph/0612571.

\bibitem[Smith et al.(2009)]{smith09} 
Smith K.M., Senatore L., and Zaldarriaga M., JCAP 0909 (2009) 006 (arXiv:0901.2572).

\bibitem[The Planck Collaboration (2005)]{bluebook} 
The Planck Collaboration, ESA-SCI(2005) [arXiv: astro-ph/0604069]

\bibitem[Tomita(1986)]{tomita86} 
Tomita H., Progr.\ Theor.\ Phys.\ 76 (1986) 952.

\bibitem[Vielva et al.(2004)]{vielva04} 
Vielva P. Mart\'inez-Gonz\'alez  E., Barreiro R.B., Sanz J.L., and Cay\'on L., \apj\ 609 (2004) 22.

\bibitem[Yadav et al.(2008)]{yadav08a} 
Yadav A.P.S., Komatsu E., Wandelt B.D., Liguori M., Hansen F.K., and Matarrese S., \apj\ 678 (2008) 578.
1
\bibitem[Yadav \& Wandelt(2008)]{yadavwan08} 
Yadav A.P.S. and Wandelt B.D., \prl\ 100 (2008) 181301.

\bibitem[Wu et al.(2001)]{wu01} 
Wu J.H.P. et al., \apjs\ 132 (2001) 1.

\end{thebibliography}
\end{document}